\documentclass[]{elsart}
\usepackage{amsmath}
\usepackage{graphicx}
\usepackage{epsfig}

\newcommand{\xm}{$X_{\rm max}$\,}
\newcommand{\xmph}{$X_{\rm max}^{\gamma, \rm med}$\,}

\newcommand{\gcm}{g\,cm$^{-2}$\,}

\def \pao {~[Pierre Auger Collaboration]}
\def \fulla {Pierre Auger Collaboration [J. Abraham et al.]}
\def \icrcmx {30th International Cosmic Ray Conference (ICRC 07), M\'{e}rida, Yucatan, Mexico (3-11 July 2007),}

\def \icrcin {29th International Cosmic Ray Conference (ICRC 05), Pune, India (3-10 August 2005),}
\def \hires {~[HiRes Collaboration]}

\begin{document}

\begin{frontmatter} 

\title{
Upper limit on the cosmic-ray photon fraction
at EeV energies from the Pierre Auger Observatory
}


\par\noindent
{\bf The Pierre Auger Collaboration} \\
J.~Abraham$^{8}$, 
P.~Abreu$^{71}$, 
M.~Aglietta$^{53}$, 
C.~Aguirre$^{12}$, 
E.J.~Ahn$^{87}$, 
D.~Allard$^{30}$, 
I.~Allekotte$^{1}$, 
J.~Allen$^{90}$, 
P.~Allison$^{92}$, 
J.~Alvarez-Mu\~{n}iz$^{78}$
% ...

\par\noindent
$^{1}$ Centro At\'{o}mico Bariloche and Instituto Balseiro (CNEA-
UNCuyo-CONICET), San Carlos de Bariloche, Argentina \\
% ...
$^{8}$ Observatorio Meteorologico Parque Gral.\ San Martin (UTN-
FRM/CONICET/CNEA), Mendoza, Argentina \\
% ...
$^{12}$ Universidad Catolica de Bolivia, La Paz, Bolivia \\
% ...
$^{30}$ Laboratoire AstroParticule et Cosmologie (APC), 
Universit\'{e} Paris 7, CNRS-IN2P3, Paris, France \\
% ...
$^{53}$ Istituto di Fisica dello Spazio Interplanetario (INAF),
 Universit\`{a} di Torino and Sezione INFN, Torino, Italy \\
% ...
$^{71}$ LIP and Instituto Superior T\'{e}cnico, Lisboa, Portugal \\
% ...
$^{78}$ Universidad de Santiago de Compostela, Spain \\
% ...
$^{87}$ Fermilab, Batavia, IL, USA \\
% ...
$^{90}$ New York University, New York, NY, USA \\
% ...
$^{92}$ Ohio State University, Columbus, OH, USA \\
% ...

\begin{abstract}
From direct observations of the longitudinal development of ultra-high
energy air showers performed with the Pierre Auger Observatory,
upper limits of 3.8\%, 2.4\%, 3.5\% and 11.7\% (at 95\% c.l.) are obtained 
on the fraction of cosmic-ray photons above 2, 3, 5 and 10~EeV 
(1~EeV $\equiv 10^{18}$ eV) respectively. 
These are the first experimental limits on ultra-high energy photons at 
energies below 10~EeV.
The results complement previous constraints on top-down models from array 
data and they reduce systematic uncertainties in the interpretation of 
shower data in terms of primary flux, nuclear composition and proton-air 
cross-section.

\end{abstract}

\end{frontmatter} 


\section{Introduction}

Data taken at the Pierre Auger Observatory were searched previously for
ultra-high energy (UHE) photons above 10~EeV~\cite{hyb_limit,sd_limit}.
In Ref.~\cite{hyb_limit}, the depth of shower maximum \xm of air
showers observed
by fluorescence telescopes in hybrid mode (i.e.\ with additional
timing information from the ground array) was used to place an 
upper limit of 16\% on the photon fraction above 10~EeV, confirming 
and improving on previous limits from ground 
arrays~\cite{agasa,risse05,ave,ave02}.
In Ref.~\cite{sd_limit}, the larger number of events taken with
the Auger ground array alone allowed us to place a limit of 2\% 
above 10~EeV, which imposes severe constraints on ``top-down'' models 
for the origin of ultra-high energy cosmic rays.

Observations in hybrid mode are also possible at energies
below 10~EeV. Decreasing the energy threshold increases the
event statistics, which to some extent balances the
factor $\sim$10 smaller duty cycle compared to observations with
the ground array alone.
Thus, based on the previous work, the search for photons is
now extended to lower energy (here down to 2~EeV).
We also improve on our previous (statistics-limited) bound
above 10~EeV from Ref.~\cite{hyb_limit}.

Photons at EeV energies are expected to be produced
in our cosmological neighborhood, as the energy attenuation length of
such photons is only of the order of a few Mpc.
Possible sources of EeV photons are the standard GZK process 
(see e.g. Refs.~\cite{GZKphot07,GZKphot,sigl}), the production by nuclei 
in regions of intense star light (e.g.\ in the galactic center~\cite{kusenko}), 
or exotic scenarios such as top-down models (see Ref.~\cite{revsigl} 
for a review). 
Compared to our previous constraints on top-down models
from Ref.~\cite{sd_limit},  the bounds derived in this work provide a test
of model predictions in a different energy range and using a different 
experimental technique, thus giving an independent confirmation of the 
model constraints.

Limits on EeV photons reduce corresponding systematic
uncertainties in other analyses of air shower data.
For instance, the presence of a substantial photon component
can severely affect the reconstruction of the energy spectrum~\cite{busca}, 
the derivation of the proton-air cross-section~\cite{belov,ulrich}, and the 
interpretation of the observed average \xm\cite{unger} in terms of a 
nuclear primary composition.

The structure of the paper is as follows. 
In Section~\ref{sec_data} the analysis 
is described and applied to the data. 
The results are discussed in Section~\ref{sec_discussion}.

\section{Data and Analysis}
\label{sec_data}

The present analysis follows closely the one described in detail
in Ref.~\cite{hyb_limit} which is called {\it Hybrid-1} below.
The basic idea is to compare the measured \xm values to those 
expected for primary photons, because UHE photon showers have 
significantly deeper average \xm. 
We provide a summary of the analysis method, paying special 
attention to differences or changes in the approach compared to 
{\it Hybrid-1}.

The data used here were taken with a total of 18 fluorescence telescopes
located at three sites (``Los Leones'',``Los Morados'' and ``Coihueco'') between
1 December 2004 and 31 December 2007. The number of ground
stations grew in this period from about 530 to 1450. 
Compared to {\it Hybrid-1} the data set above 10~EeV increased in size 
by a factor $\sim$2.2.

The event reconstruction~\cite{offline} is based on an end-to-end
calibration of the fluorescence telescopes~\cite{knapik}, monthly models for the
atmosphere~\cite{keilhauer}, and an average aerosol model based on local 
atmospheric measurements~\cite{aerosol}.
The reconstruction of the longitudinal profile is described 
in~\cite{unger_prof}.
A correction of $\sim$~1\% for the missing energy (energy carried by neutrinos 
or high-energy muons) is applied to the reconstructed calorimetric energy, 
corresponding to the effective energy of primary photons~\cite{pierog}. 

The following quality cuts are applied to the collected events:
\begin{itemize}
\item[$\bullet$] number of phototubes in the fluorescence telescope 
triggered by the shower $\ge$6;
\item[$\bullet$] distance of closest approach of the reconstructed shower
 axis to the surface detector station with the largest signal is $<$1.5~km, 
 and difference between the reconstructed shower front arrival time
 at this station and the measured tank time is $<$300~ns;
\item[$\bullet$] normalized $\chi^2_{\rm prof}$ of the longitudinal shower 
profile fit~\cite{unger_prof} $<$6,
and ratio of $\chi^2_{\rm prof}$ to $ \chi^{2}_{\rm line}< 0.9$, 
where $\chi^{2}_{\rm line}$
 refers to a straight line fit (the latter cut essentially rejects 
profiles with too few data points);
\item[$\bullet$] depth of shower maximum \xm observed
 in the telescope field of view (this cut may be relaxed in future 
to allow also the search for deeply penetrating events with \xm beyond 
the field of view);
\item[$\bullet$] minimum angle between the viewing direction of a 
triggered pixel and the shower axis $>$15$^\circ$ 
(to reject events with a large Cherenkov light contamination);
\item[$\bullet$] primary energy $E > f \cdot$~EeV,
 $f = 2,~3,~5,~10$ (the analysis in {\it Hybrid-1} was restricted to
 $f=10$).
\end{itemize}

The criterion of \xm being observed can introduce
a bias against the deeply penetrating photon primaries (e.g. for
near-vertical events). To reduce the dependence of the detector acceptance 
on composition, fiducial volume cuts are applied:
%
%
\begin{itemize}
\item shower zenith angle~$>$~35$^\circ~+~g_1(E)$
$$ g_{1}(E)= 
\begin{cases}
10^{\,\circ}~(\lg E/\rm{eV}-19.0) & \text{for $\lg E/$eV~$\le19.7$,}\\
7^{\,\circ} & \text{for $\lg E/$eV~$>$~19.7;}
\end{cases}
$$
\item distance of telescope to shower core~$<$~24~km~$+~g_2(E)$
$$ g_{2}(E)= 
\begin{cases}
12~(\lg E/\rm{eV}-19.0)~km & \text{for $\lg E/$eV~$\ge19.0$,}\\
6~(\lg E/\rm{eV}-19.0)~km & \text{for $\lg E/$eV$<19.0$.}
\end{cases}
$$
\end{itemize}
The described cuts are identical to those from {\it Hybrid-1} for showers
$>$10~EeV, but allow now for an extension of the energy range down to 2~EeV.

\begin{figure}[t]
\begin{center}
\includegraphics[height=6.1cm]{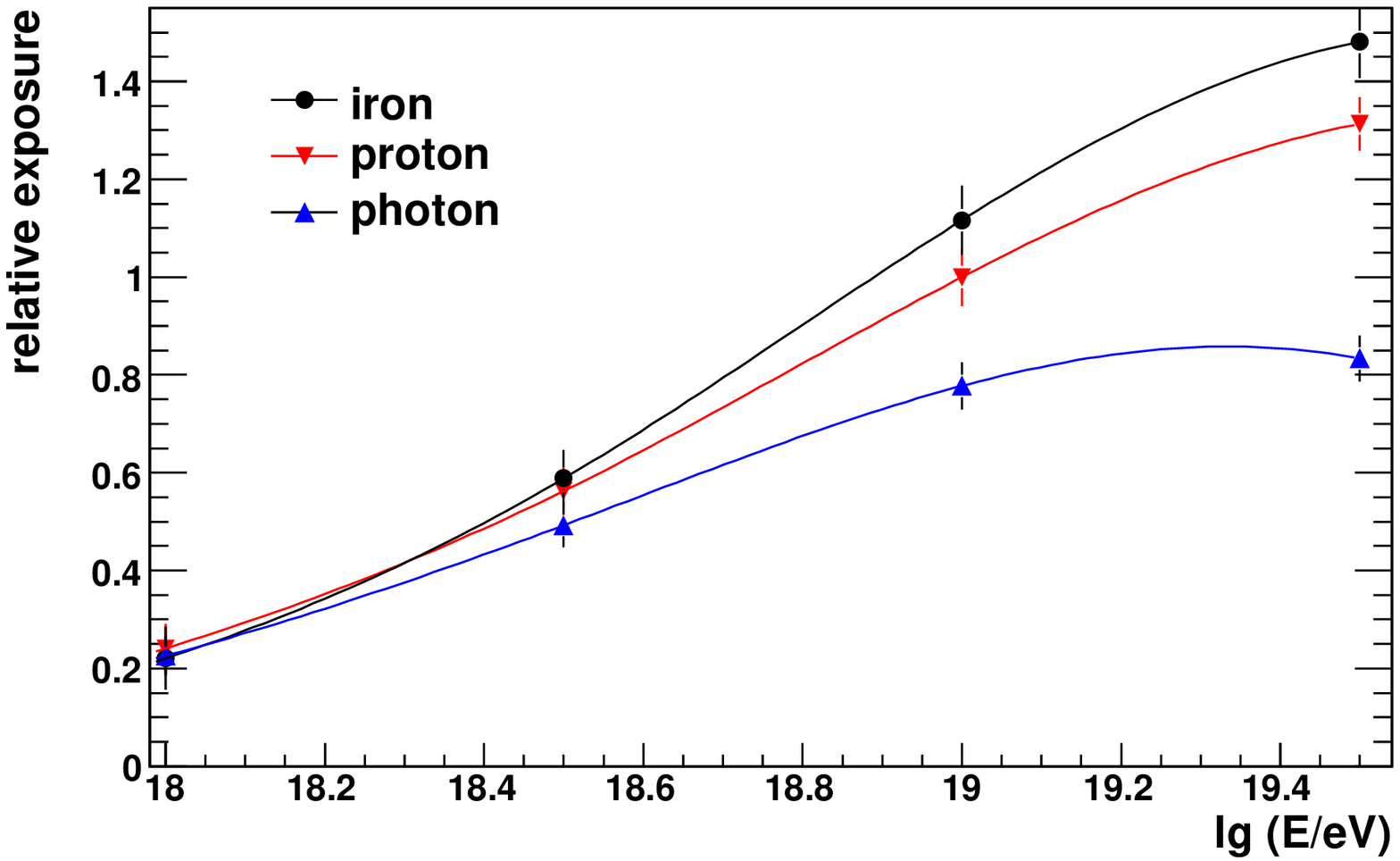}
\includegraphics[height=6.1cm]{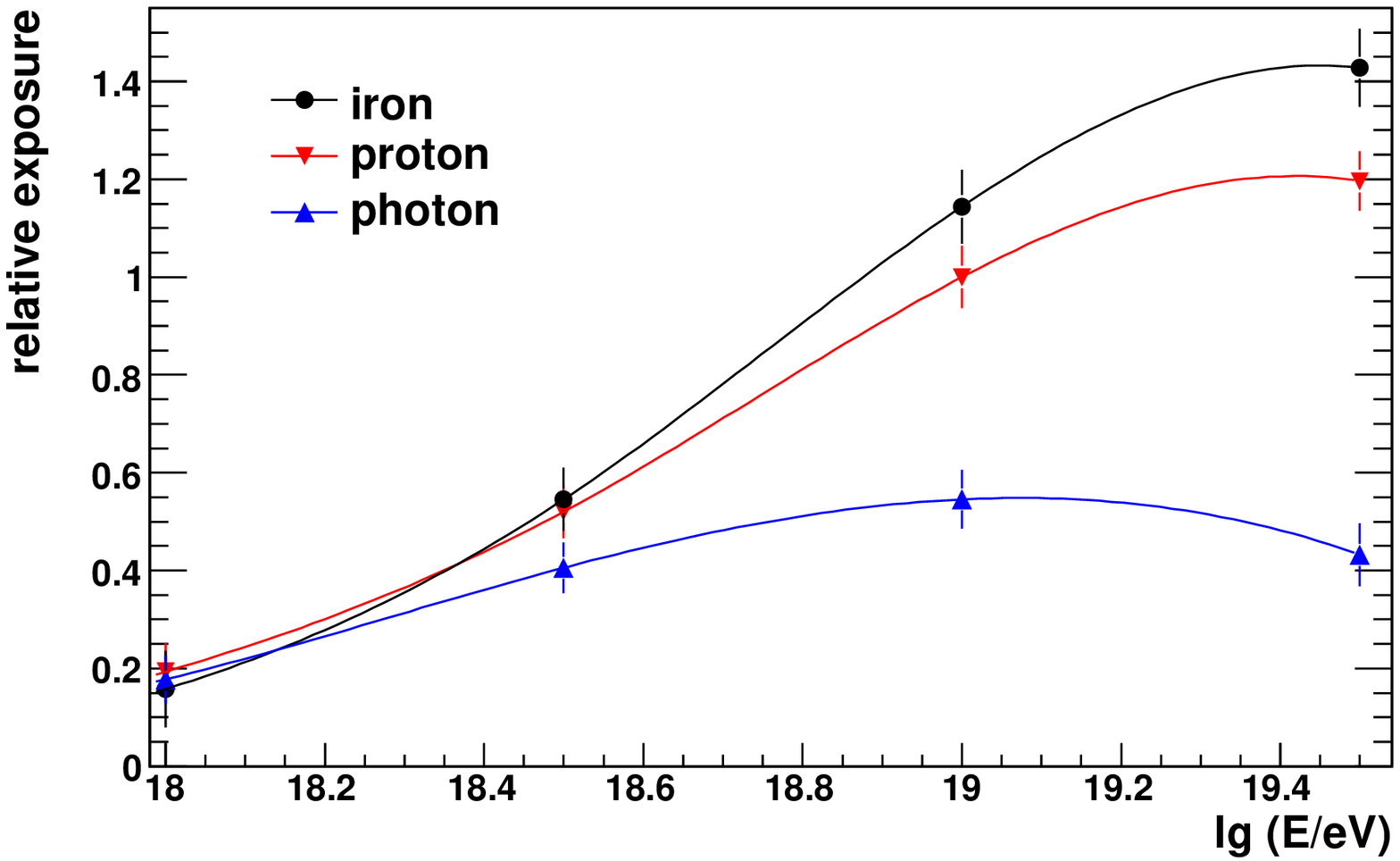}
\includegraphics[height=6.1cm]{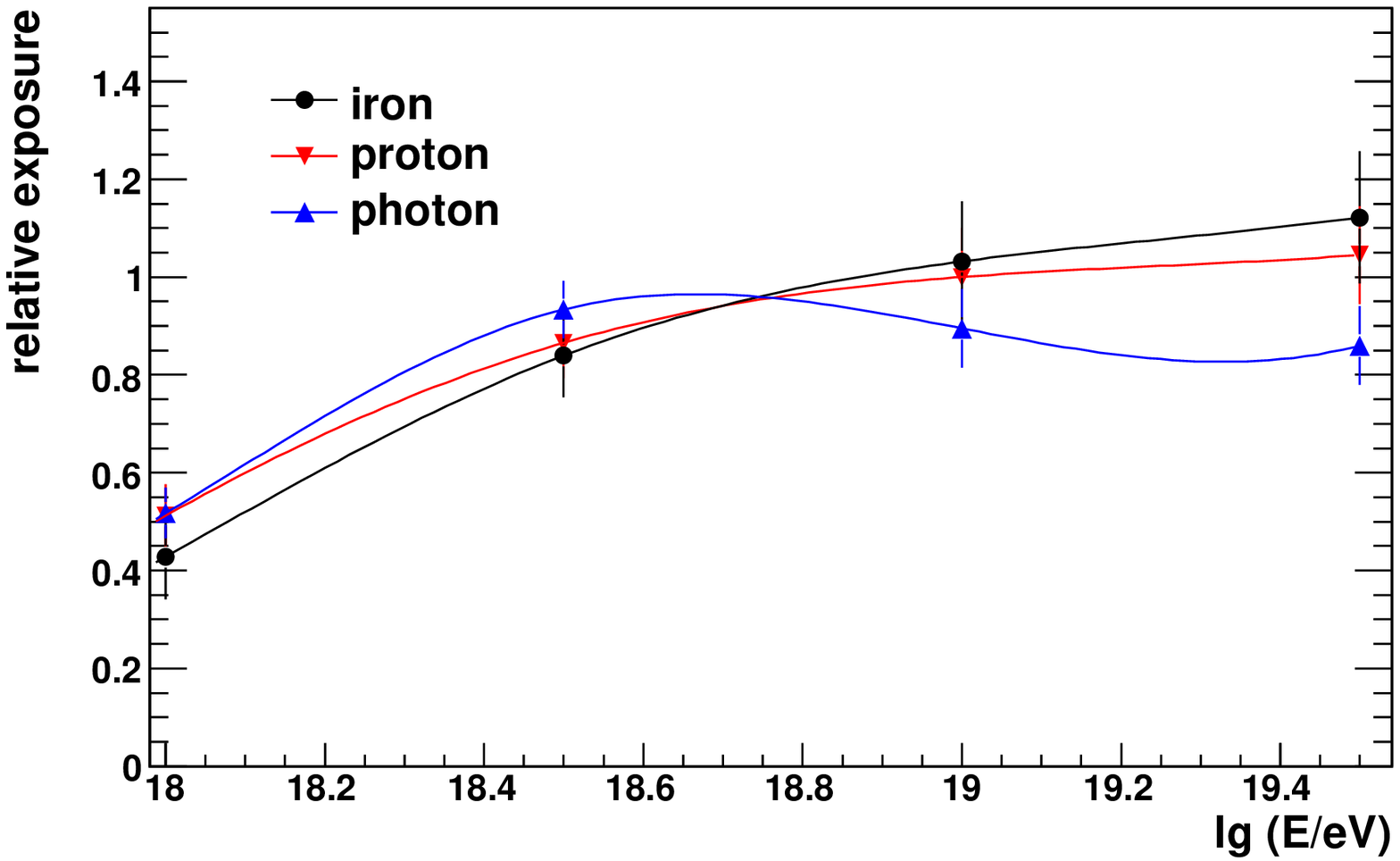}
\caption{Relative exposure to primary photons, protons and
iron nuclei, normalized
to protons at 10~EeV. Top panel requiring hybrid trigger, center
panel after applying quality cuts, bottom panel after applying 
fiducial volume cuts (see text). In order to guide the eye 
polynomial fits are superimposed to the obtained values.
\label{fig-eff}}
\end{center}
\end{figure}

To evaluate the detector acceptance as a function of energy for different 
primary particles, simulations have been performed using 
CORSIKA~\cite{corsika} with QGSJET01~\cite{qgsjet} and FLUKA~\cite{fluka} 
as high- and low-energy hadronic interaction models respectively. 
The Monte Carlo showers have been processed through a complete detector 
simulation and reconstruction chain~\cite{offline,fdsim}. 
In Fig.~\ref{fig-eff} we show the energy-dependent relative exposure 
obtained after trigger, quality cuts, and fiducial volume cuts for 
primary photons, protons and iron nuclei (normalized to 10~EeV protons).
After fiducial volume cuts, the acceptance for photons is close to
 the acceptance for nuclear primaries.
Thus, the relative abundances of photon and nuclear
primaries are preserved to a good approximation. In a similar way 
to {\it Hybrid-1},
we apply, for the derivation of an upper limit on the photon fraction,
an efficiency correction according to the acceptances after fiducial 
volume cuts which is conservative and independent of assumptions about 
the actual primary fluxes (factor ``$\epsilon_{\rm fvc}$'', see Appendix).

Applying the selection cuts to the data, there remain
$n_{\rm total}^\prime (E_{\rm thr}^\gamma) =$
2063, 1021, 436 and 131 events with energies greater than 
$E_{\rm thr}^\gamma =$
2, 3, 5 and 10~EeV respectively.
The label $\gamma$ in $E_{\rm thr}^\gamma$ indicates that the
missing energy correction for photons has been applied.
To obtain $n_{\rm total} (E_{\rm thr}^\gamma)$ from the total number 
of events $n_{\rm total}^\prime (E_{\rm thr}^\gamma)$ after fiducial 
volume cuts, those events need to be rejected where clouds may have 
disturbed the observation. 
The presence of clouds could change the efficiencies which are shown in 
Fig.~\ref{fig-eff}. Also, the reconstructed \xm values may be affected. 
Particularly, clouds may obscure early parts of the shower profile such 
that the remaining event profile looks deeply penetrating and, hence, 
photon-like. Therefore we only use data where any disturbance by clouds 
can be excluded using information from the IR cloud monitoring 
cameras~\cite{cester,clffick}.
In {\it Hybrid-1} all events were individually checked.
As this is hardly feasible for the events in the present data set
(a full automatic processing of cloud data is in preparation), the 
following approach is adopted. To determine the efficiency 
$\epsilon_{\rm clc}$ of passing the {\it cl}oud {\it c}ut we used 
the sample of events with energy above 10~EeV. Accepting only events 
where any disturbance by clouds could be excluded, 67 events out of 
131 have been selected, corresponding to $\epsilon_{\rm clc} \simeq$ 0.51. 
We confirmed that this efficiency also 
holds at lower energy by applying the same criteria to a sub-set of 
$\sim$300 events at $\sim$3~EeV.
The final number of $n_{\rm total} (E_{\rm thr}^\gamma)$
is then given by 
$n_{\rm total} (E_{\rm thr}^\gamma)
  = \epsilon_{\rm clc} \cdot n_{\rm total}^{\prime}(E_{\rm thr}^\gamma)$.

As the present data set above 2~EeV is about a factor $\sim$15 larger
than the one used in {\it Hybrid-1}, a different statistical method 
is applied to derive the photon limit. 
For the derivation of the limit in {\it Hybrid-1}, each selected event 
was individually compared with high-statistics photon simulation, using 
the respective primary energy and direction as simulation input.
This method is CPU demanding, and tailormade for a relatively small 
number of events. 
We therefore adopt for our analysis the method applied in Ref.~\cite{sd_limit}
which needs as an input the total number of events, the number of 
photon candidates (events having ``photon-like'' characteristics, see below) 
and proper correction factors accounting for inefficiencies. 
The 95\% c.l.\ upper limit $F_{\gamma}^{\,95} (E_{\rm thr})$ on 
the fraction of photons in the cosmic-ray flux above $E_{\rm thr}$
is then given by

\begin{equation}
\label{eq:pfrac1}
  F_{\gamma}^{\,95} (E_{\rm thr})
   = \frac{ n_{\gamma-{\rm cand}}^{95} (E_{\rm thr}^\gamma) }
          { n_{\rm total} (E_{\rm thr}^\gamma) }~,
\end{equation}

where $n_{\gamma-{\rm cand}}^{95}$ is the 95\% c.l.\ upper limit
on the number of photon candidates and $n_{\rm total}$ the total
number of selected events. 
As it is not known in advance whether photons indeed compose 
only a negligible fraction of the cosmic-ray flux, we apply the missing 
energy correction appropriate for photons to all events and take here 
$n_{\rm total} (E_{\rm thr}^\gamma)$.
This is conservative (larger value of $ F_{\gamma}^{\,95}$), since using the 
missing energy correction for hadrons (factor $\simeq 1.07-1.14$ \cite{barbosa,pierog}) would 
increase the total number of events above $E_{\rm thr}$, i.e.\
$n_{\rm total} (E_{\rm thr}^\gamma) 
 < n_{\rm total} (E_{\rm thr}^{\rm had})$. 
%
\begin{figure}[t!]
\begin{center}
\includegraphics[height=9.0cm]{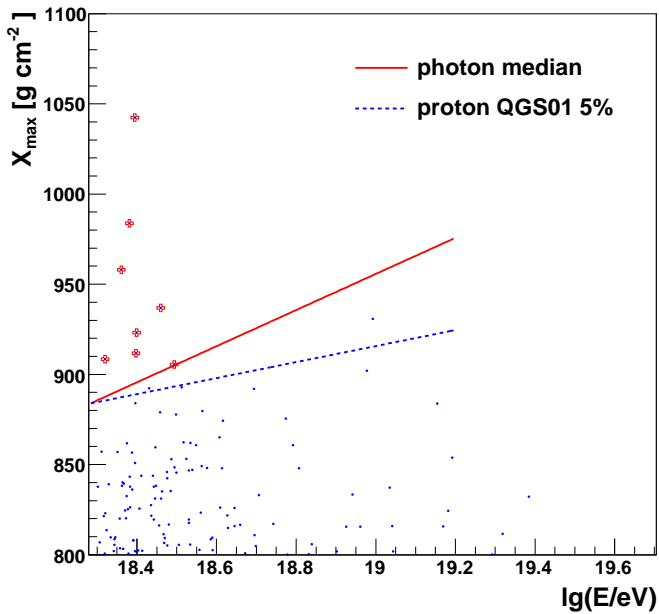}
\caption{Closeup of the scatter plot of \xm vs. energy for all events
(blue dots) with \xm above 800~\gcm and energy above 2~EeV, after
quality, fiducial volume and cloud cuts.
Red crosses show the 8 photon candidate events (see text). 
The solid red line indicates the typical median depth of
shower maximum for primary photons, parameterized as \xmph$=a\cdot y +b$, 
for $y=\lg(E/\rm{EeV})$, $y=[0,1.2]$, where $a=100$~\gcm and $b=856$~\gcm.
The dashed blue line results from simulations of primary protons using 
QGSJET 01. A fraction of 5\% of the simulated proton showers had \xm 
values larger than indicated by the line.
\label{fig-xmaxdistr}}
\end{center}
\end{figure}

A scatter plot of \xm vs. energy for all events above 
$E_{\rm thr}^\gamma$=2~EeV with \xm $\ge 800$~\gcm surviving quality, fiducial 
volume and cloud cuts is shown in Fig.~\ref{fig-xmaxdistr}. Statistical 
uncertainties in individual events are typically a few percent in energy 
and $\sim 15-30$~\gcm in \xm. Systematic uncertainties are $\sim 22$\% 
in energy~\cite{spectrumPRL} and $\sim11$~\gcm in \xm~\cite{unger}. 

The upper limit on the number of photon candidates
$n_{\gamma-{\rm cand}}^{95}$ is given by
$n_{\gamma-{\rm cand}}^{95} =
  n_{\gamma-{\rm cand, obs}}^{95} / \epsilon_{\rm obs}$,
where $n_{\gamma-{\rm cand, obs}}^{95}$ is the
95\% c.l.\ upper limit on the number of photon candidates
$n_{\gamma-{\rm cand, obs}}$ 
extracted (``observed'') from the data set and
$\epsilon_{\rm obs}$ is the corresponding efficiency.
$n_{\gamma-{\rm cand, obs}}$ is taken as the number of
events which have the observed \xm above the median
\xmph of the distribution expected for photons
of that energy and direction (``photon candidate cut'').
Additionally, on these particular events individual cloud 
checks have been performed, and only events that pass this 
cloud check are finally considered as photon candidates. 
In Fig.~\ref{fig-xmaxdistr}, 
typical values of \xmph$(E)$ are indicated as a function of energy (solid red line). 
To extract the specific value of \xmph for each individual event, 
dedicated simulations with primary photons have been performed for all 
potential candidate events, assuming the corresponding energy and geometry.

\begin{figure}[t]
\begin{center}
\includegraphics[height=5.4cm]{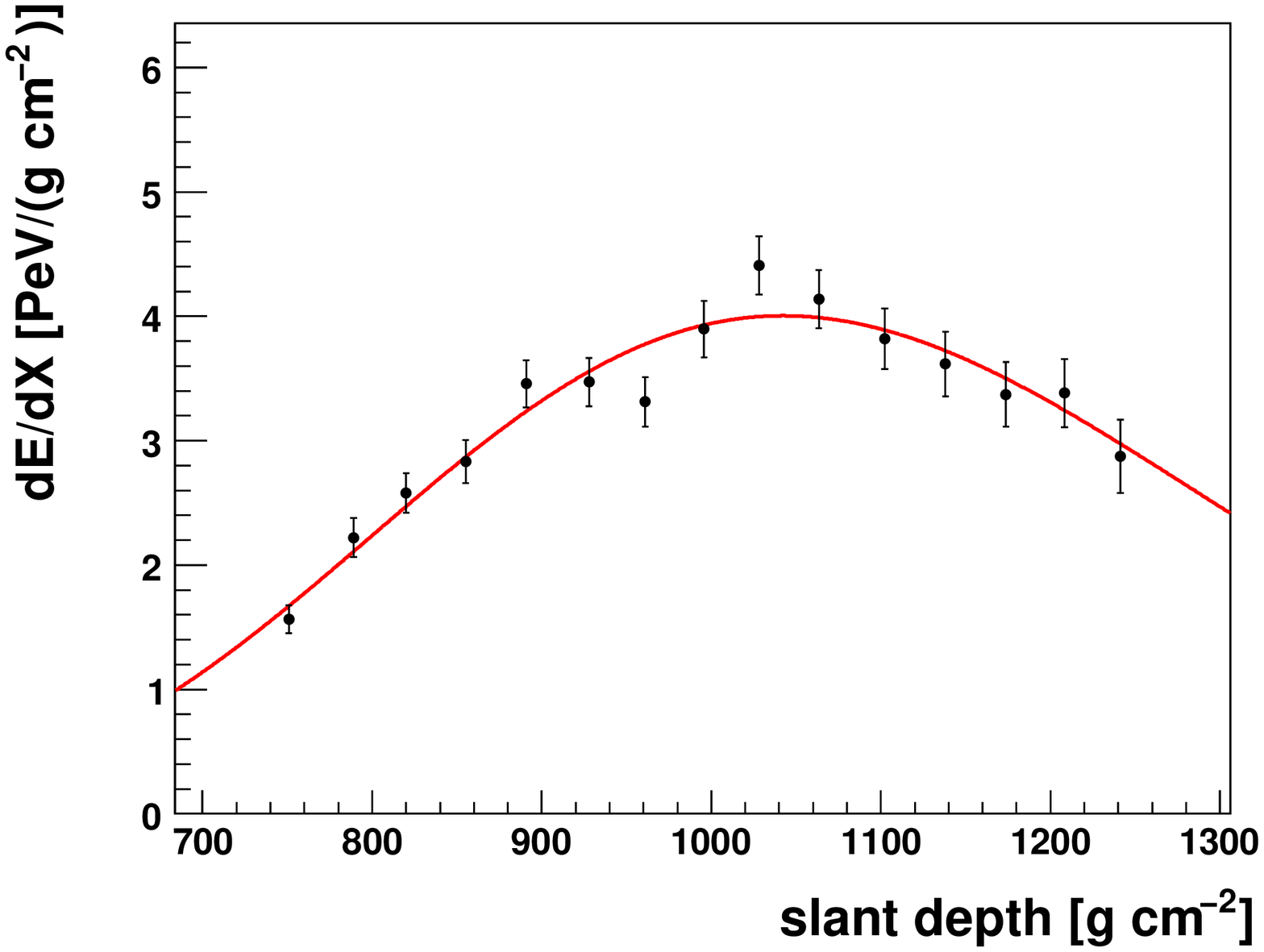}
\hspace{0.25cm}
\includegraphics[height=5.8cm]{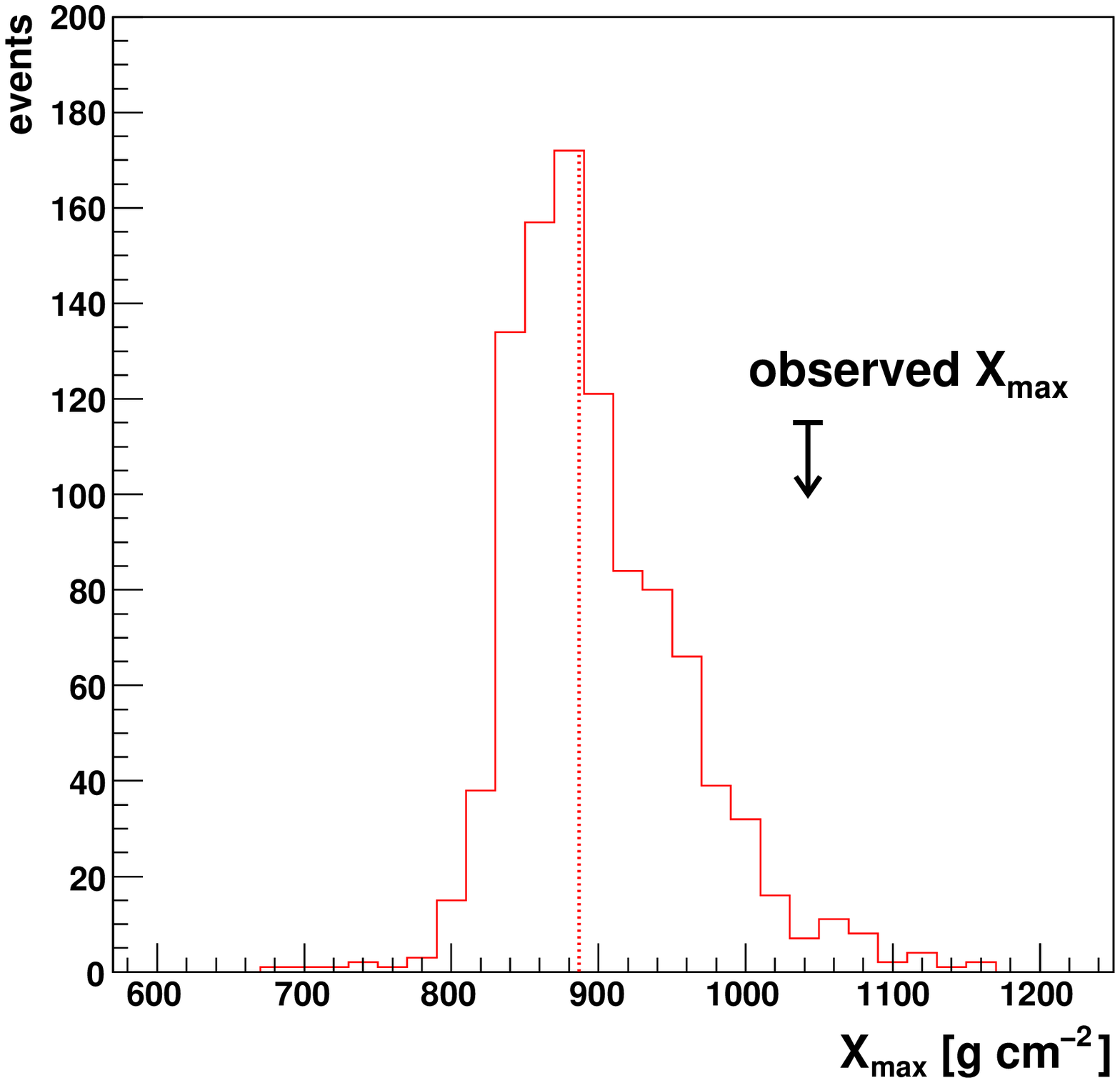}
\caption{Left panel: shower profile (black bullets) of the 
deepest \xm candidate event in the analyzed sample (id 3554364), 
along with the Gaisser-Hillas fit (red line). 
Right panel: the observed \xm value (black arrow) along with the 
\xm distribution from the dedicated photon simulation (histogram); 
see Tab.~\ref{tab:phcandy} for statistical uncertainty. 
The dashed line indicates 
the median of the photon distribution.
\label{fig-event}}
\end{center}
\end{figure}

There are $n_{\gamma-{\rm cand, obs}}$~=~8, 1, 0, 0 photon candidate 
events with energies greater than 2, 3, 5 and 10~EeV, respectively.
These candidate events are marked by red crosses in Fig.~\ref{fig-xmaxdistr}
and the event parameters are listed in Table~\ref{tab:phcandy}.
As an illustration, the shower profile of the candidate with the 
deepest \xm is displayed in the left panel of Fig.~\ref{fig-event};
in the right panel the measured \xm value is shown along with the
results of the dedicated photon simulations.

\begin{table}[b!]
\begin{center}
\caption{Characteristic parameters for the eight events 
surviving the photon candidate cut ($\Delta$\xm refers to 
the statistical uncertainty).}
\label{tab:phcandy}
\vskip 0.2cm
\newcommand{\m}{\hphantom{$-$}}
\newcommand{\cc}[1]{\multicolumn{1}{c}{#1}}
\renewcommand{\tabcolsep}{0.7pc} 
\renewcommand{\arraystretch}{1.2} 
\begin{tabular}{cccc}
\hline
 {id} & {\bf \xm} [\gcm] &{\bf $\Delta$\xm} [\gcm] & {\bf $E_{\gamma}$} [EeV] 
\\ \hline
2051232 & 923 &  17 &  2.5 \\
2053796 & 905 &  32 &  3.1 \\
2201129 & 958 &  29 &  2.3 \\
2566058 & 908 &  20 &  2.1 \\
2798252 & 937 &  29 &  2.9 \\
3478238 & 984 &  12 &  2.4 \\
3554364 & 1042 & 12 &  2.5 \\
3690306 & 912 &  27 &  2.5 \\ 
\hline
\end{tabular}\\[2pt]
\end{center}
\end{table}

We checked with simulations whether the observed number of photon candidate 
events is significantly larger than the expectation in case of nuclear 
primaries only, i.e. whether primary photons appear to be required to 
explain the photon candidates. The quantitative estimation of the 
background expected from nuclear primaries suffers from substantial 
uncertainties, namely the uncertainty of the primary composition in this 
energy range (a larger background to photons would originate from 
lighter nuclear primaries) and the uncertainty in the high-energy 
hadronic interactions models (for instance, reducing the proton-air 
cross-section allows proton primaries to penetrate deeper into the 
atmosphere). From simulations using QGSJET01 as the hadronic interaction 
model, we found that the observed number of photon candidate
events is well within the number of background events expected from a 
pure proton and a pure iron composition. For energies larger than 
2~EeV about 30 events are expected in the analyzed time window for 
proton and 0.3 for iron. The corresponding numbers above 3, 5, 10~EeV 
are about 12, 4, 1 events for proton and about 0.2, 0.1, 0.0 events 
for iron. Scenarios of a mixed composition, as also favored by our 
results on $<$\xm$>$~\cite{unger}, can reproduce the observation. 
We conclude that the observed photon candidate events may well be due 
to nuclear primaries only. This also holds for the candidate event with 
the largest \xm shown in Fig.~\ref{fig-event}: proton showers with
comparable or larger \xm value occur at a level of a few out of
thousand simulated events. 

We now continue to derive the upper limit to the photon fraction. 
$n_{\gamma-{\rm cand, obs}}^{95}$ is calculated from $n_{\gamma-{\rm cand, obs}}$ 
using the Poisson distribution and assuming no background,
i.e.\ $n_{\gamma-{\rm cand, obs}}$ is not reduced by
subtracting any event that may actually be due to nuclear primaries. 
This procedure represents the most conservative approach as it 
maximizes the value of $n_{\gamma-{\rm cand, obs}}^{95}$. 
The efficiency $\epsilon_{\rm obs}$ of photons passing all cuts
is given by
$\epsilon_{\rm obs} = \epsilon_{\rm fvc} \epsilon_{\rm pcc}$ 
where $\epsilon_{\rm fvc} \simeq 0.72 - 0.77$ 
(see Tab.~\ref{tab:phlimits}) comes from the acceptance after 
{\it f}iducial {\it v}olume {\it c}uts (see Appendix) and, by construction, 
$\epsilon_{\rm pcc} = 0.50$ is given by the {\it p}hoton {\it c}andidate 
{\it c}ut above the median of the \xm 
distribution for photons. Thus, the upper limit is calculated according to

\begin{equation}
\label{eq:pfrac2}
  F_{\gamma}^{\,95} (E_{\rm thr})
   = \frac{
       n_{\gamma-{\rm cand, obs}}^{95} (E_{\rm thr}^\gamma)
             ~\frac{1}{ \epsilon_{\rm fvc}}
            ~ \frac{1}{ \epsilon_{\rm pcc}} }
          { n_{\rm total}^\prime (E_{\rm thr}^\gamma) ~ \epsilon_{\rm clc} }~.
\end{equation}

Applied to the data, upper limits of 3.8\%, 2.4\%, 3.5\% and 11.7\% on 
the fraction of cosmic-ray photons above 2, 3, 5 and 10~EeV
are obtained at 95\% c.l.. Table~\ref{tab:phlimits} provides a summary 
of the quantities used in the derivation of the integral upper limits. 
\begin{table}[hbt]
\begin{center}
\caption{Summary of the quantities used in the derivation of the 
integral upper limits on the photon fraction for 
$E_{\rm thr}^\gamma$~=~ 2, 3, 5, and 10~EeV. Not listed are the 
efficiencies  $\epsilon_{\rm clc} = 0.51$ and $\epsilon_{\rm pcc} = 0.50$ 
which do not depend on $E_{\rm thr}^\gamma$.}
\vskip 0.2cm
\label{tab:phlimits}
\newcommand{\m}{\hphantom{$-$}}
\newcommand{\cc}[1]{\multicolumn{1}{c}{#1}}
\renewcommand{\tabcolsep}{0.9pc} 
\renewcommand{\arraystretch}{1.3} 
\begin{tabular}{lccccc}
\hline
$E_{\rm thr}^\gamma$ [EeV] &$n_{\gamma-{\rm cand, obs}}$ &$n_{\gamma-{\rm cand, obs}}^{95}$ &  $n_{\rm total}^\prime$ & $\epsilon_{\rm fvc}$&  $F_{\gamma}^{\,95}$[\%] 
\\ \hline
2    &  8& 14.44 &  2063 & 0.72 & 3.8\\
3    &  1& 4.75  &  1021 & 0.77 & 2.4\\
5    &  0& 3.0   &  436  & 0.77 & 3.5\\
10   &  0& 3.0   &  131  & 0.77 & 11.7\\
\hline
\end{tabular}\\[2pt]
\end{center}
\end{table}

We studied the robustness of the results against different sources
of uncertainty. 
Varying individual event parameters or the selection criteria, within
the experimental resolution, leaves the results essentially unchanged.
Uncertainties in the determination of the efficiency factors
used in Eq.~\ref{eq:pfrac2}
are estimated to correspond to an uncertainty
$\Delta F_{\gamma}^{95} / F_{\gamma}^{95} \simeq 0.15$.
Increasing (reducing) {\it all} reconstructed \xm values
by $\Delta X_{\rm max}^{\rm syst} = 11$~\gcm\cite{unger} 
changes the number of photon candidates above 2~EeV by
$+$1 ($\pm$0) and above 3~EeV by $\pm$0 ($-$1), while it does not 
affect the higher energies.
The limits then become 4.1\% (3.8\%) above 2~EeV and 2.4\% (1.5\%) 
above 3~EeV. The energy scale $E_{\rm thr}$ which the limit
$F_{\gamma}^{95} (E_{\rm thr})$ refers to, has a 22\%
systematic uncertainty~\cite{spectrumPRL}. Hence, the numerical values of the 
limits $F_{\gamma}^{95}$ derived here refer to an effective energy 
threshold
$E_{\rm thr}^{\rm eff} = k_E \times E_{\rm thr}$,
with $k_E = 0.78... 1.22$.
Related to an increase (reduction) of the energy scale is a small
upward (downward) shift of the \xm value used for the
photon candidate cut, leading to stronger (weaker) criteria for
an event to pass this cut.
This shift amounts to $\sim$7~\gcm for a 22\% change of
the energy scale.
Finally, an uncertainty $<$10~\gcm on the simulated
photon \xm values comes from the need to extrapolate
the photonuclear cross-section to high energy~\cite{risse06}.
Adding in quadrature the discussed uncertainties in \xm 
gives an effective total uncertainty of $\sim$16~\gcm.
Increasing (reducing) {\it all} reconstructed \xm values
by this amount changes the number of photon candidates above
2 and 3~EeV by $+$3 ($\pm$0) and by $+$1 ($-$1). 
Accordingly the limits then become 4.8\% (3.8\%) above 2~EeV 
and 3.1\% (1.5\%) above 3~EeV, while the limits above 5 and 10~EeV 
are unchanged.

\begin{figure}[t!]
\begin{center}
\includegraphics[height=8.4cm]{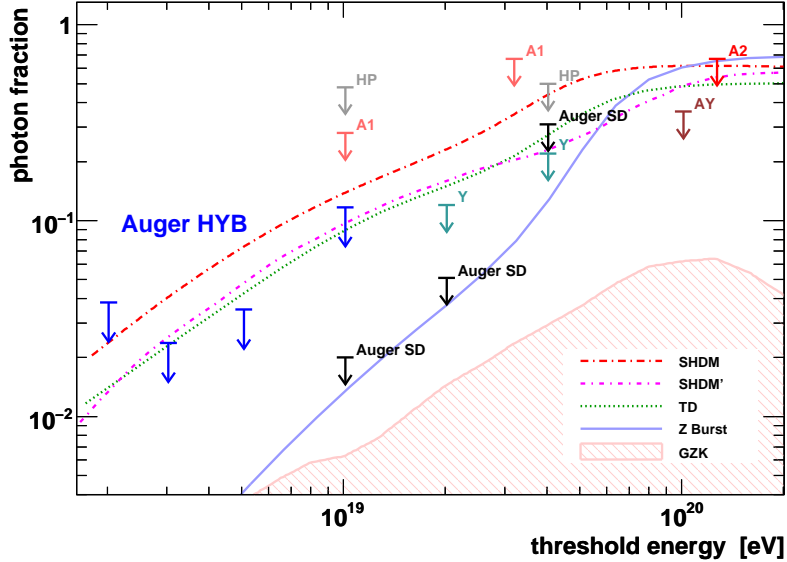}
\caption{Upper limits on the photon fraction in the integral cosmic-ray 
flux for different experiments: AGASA (A1, A2)~\cite{agasa,risse05}, 
AGASA-Yakutsk (AY)~\cite{agayak}, Yakutsk (Y)~\cite{yakutsk}, 
Haverah Park (HP)~\cite{ave,ave02}. 
In black the limits from the Auger surface detector (Auger SD)~\cite{sd_limit} 
and in blue the limits above 2, 3, 5, and 10~EeV derived in this work 
(Auger HYB). The shaded region shows the expected GZK photon fraction 
as derived in~\cite{GZKphot07}. Lines indicate predictions from top-down 
models, see~\cite{GZKphot,shdm} and~\cite{risse07}. 
\label{fig-limits}}
\end{center}
\end{figure}
%

\section{Discussion}
\label{sec_discussion}

The derived upper limits are shown in Fig.~\ref{fig-limits}
along with previous experimental limits and model predictions 
(see Ref.~\cite{risse07} for a review and references).
These new bounds are the first ones at energies below 10~EeV 
and, together with {\it Hybrid-1},
the only ones obtained so far from fluorescence
observations (all other limits coming from ground arrays).
The results complement the previous constraints on top-down models
from Auger surface detector data.
It should be noted that due to the steep flux spectrum,
even the previous Auger bound of 2\%
above 10~EeV only marginally constrains the photon contribution
above lower threshold energies (for instance, even above
5~EeV, $\sim$75\% of the events are in the previously
untested energy range of 5$-$10~EeV).

The photon limits derived in this work also help to reduce certain 
systematic uncertainties in other analyses of air shower data such as
(i) energy spectrum: the Auger method of reconstructing the energy
spectrum does not suffer from a large contamination from photons
at EeV energies;
(ii) nuclear primary composition: the interpretation of observables
sensitive to the primary particle (for instance the observed
average \xm) in terms of a nuclear primary composition can only
be marginally biased by contributions from photons;
(iii) proton-air cross-section:
the possible contamination from photons was one of the dominant
uncertainties for deriving the proton-air cross-section~\cite{belov,ulrich},
and this uncertainty is now significantly reduced
(to $\sim$50~mb for data at EeV energies, which corresponds to a 
relative uncertainty of $\sim$10\%).

In future photon searches, the separation power between 
photons and nuclear primaries can be enhanced by adding the detailed 
information measured with the surface detectors in hybrid events. 
For an estimate of the future sensitivity of Auger to photons 
see Ref.~\cite{risse07}. The information on 
event directions can also be used in future analyses; for instance, an 
excess flux of photons from the direction of the galactic center 
(e.g. Ref.~\cite{kusenko}) can be searched for.


\textit{Acknowledgements:}

\begin{sloppypar}
The successful installation and commissioning of the Pierre Auger Observatory
would not have been possible without the strong commitment and effort
from the technical and administrative staff in Malarg\"ue.

We are very grateful to the following agencies and organizations for financial support: 
Comisi\'on Nacional de Energ\'ia At\'omica, 
Fundaci\'on Antorchas,
Gobierno De La Provincia de Mendoza, 
Municipalidad de Malarg\"ue,
NDM Holdings and Valle Las Le\~nas, in gratitude for their continuing
cooperation over land access, Argentina; 
the Australian Research Council;
Conselho Nacional de Desenvolvimento Cient\'ifico e Tecnol\'ogico (CNPq),
Financiadora de Estudos e Projetos (FINEP),
Funda\c{c}\~ao de Amparo \`a Pesquisa do Estado de Rio de Janeiro (FAPERJ),
Funda\c{c}\~ao de Amparo \`a Pesquisa do Estado de S\~ao Paulo (FAPESP),
Minist\'erio de Ci\^{e}ncia e Tecnologia (MCT), Brazil;
AVCR AV0Z10100502 and AV0Z10100522,
GAAV KJB300100801 and KJB100100904,
MSMT-CR LA08016, LC527, 1M06002, and MSM0021620859, Czech Republic;
Centre de Calcul IN2P3/CNRS, 
Centre National de la Recherche Scientifique (CNRS),
Conseil R\'egional Ile-de-France,
D\'epartement  Physique Nucl\'eaire et Corpusculaire (PNC-IN2P3/CNRS),
D\'epartement Sciences de l'Univers (SDU-INSU/CNRS), France;
Bundesministerium f\"ur Bildung und Forschung (BMBF),
Deutsche Forschungsgemeinschaft (DFG),
Finanzministerium Baden-W\"urttemberg,
Helmholtz-Gemeinschaft Deutscher Forschungszentren (HGF),
Ministerium f\"ur Wissenschaft und Forschung, Nordrhein-Westfalen,
Ministerium f\"ur Wissenschaft, Forschung und Kunst, Baden-W\"urttemberg, Germany; 
Istituto Nazionale di Fisica Nucleare (INFN),
Ministero dell'Istruzione, dell'Universit\`a e della Ricerca (MIUR), Italy;
Consejo Nacional de Ciencia y Tecnolog\'ia (CONACYT), Mexico;
Ministerie van Onderwijs, Cultuur en Wetenschap,
Nederlandse Organisatie voor Wetenschappelijk Onderzoek (NWO),
Stichting voor Fundamenteel Onderzoek der Materie (FOM), Netherlands;
Ministry of Science and Higher Education,
Grant Nos. 1 P03 D 014 30, N202 090 31/0623, and PAP/218/2006, Poland;
Funda\c{c}\~ao para a Ci\^{e}ncia e a Tecnologia, Portugal;
Ministry for Higher Education, Science, and Technology,
Slovenian Research Agency, Slovenia;
Comunidad de Madrid, 
Consejer\'ia de Educaci\'on de la Comunidad de Castilla La Mancha, 
FEDER funds, 
Ministerio de Ciencia e Innovaci\'on,
Xunta de Galicia, Spain;
Science and Technology Facilities Council, United Kingdom;
Department of Energy, Contract No. DE-AC02-07CH11359,
National Science Foundation, Grant No. 0450696,
The Grainger Foundation USA; 
ALFA-EC / HELEN,
European Union 6th Framework Program,
Grant No. MEIF-CT-2005-025057, 
and UNESCO.
\end{sloppypar}

\appendix

\section{Acceptance correction}
\label{app-acc}

The fraction of photons $f_\gamma$ in the cosmic-ray flux
integrated above an energy threshold $E_{\rm thr}$ is given by
\begin{equation}
f_\gamma(E\ge E_{\rm thr}) =
\frac{\int_{_{E_{\rm thr}}} \Phi_\gamma (E) dE}
{\int_{E_{\rm thr}} \Phi_\gamma (E) dE
+\sum_i\int_{E_{\rm thr}} \Phi_i (E) dE}
\end{equation}
where $\Phi_\gamma (E)$ denotes the differential flux of photons
and $\Phi_i (E),~ i={\rm p,He,...}$ the fluxes of nuclear primaries.

The fraction of photons $f_\gamma^{\rm det}$ as registered by
the detector is given by
\begin{equation}
f_\gamma^{\rm det}(E\ge E_{\rm thr}) =
\frac{
\int_{E_{\rm thr}} A_\gamma (E) \Phi_\gamma (E) dE 
}{
\int_{E_{\rm thr}} A_\gamma (E) \Phi_\gamma (E) dE
+\sum_i\int_{E_i} A_i (E) \Phi_i (E) dE
}
\end{equation}
with $A_\gamma (E)$ and $A_i (E)$ being the detector acceptances
to photons and nuclear primaries, respectively. $E_i$ denotes
the effective threshold energy for primary nucleus $i$.

Thus, the upper limit $f_\gamma^{\rm ul,det}$
obtained to the registered data,
$f_\gamma^{\rm ul,det} > f_\gamma^{\rm det}$, needs to be
corrected to resemble an upper limit on the fraction of
photons in the cosmic-ray flux. For the present analysis, a conservative
and model-independent correction is applied as follows.
The approach adopted here extends the one introduced in {\it Hybrid-1},
as we now also treat the case of $A_\gamma (E) \ne$ const. 

$E_{\rm thr}$ corresponds to the analysis threshold energy assuming 
primary photons. $E_i$ is related to $E_{\rm thr}$ by the ratios of 
the missing energy corrections $m_\gamma$ (for photons) and $m_i$ 
(for nuclear primaries),
\begin{equation}
E_i = E_{\rm thr} \cdot \frac{m_i}{m_\gamma}~.
\end{equation}
Since $m_\gamma \simeq 1.01$~\cite{pierog} and $m_i\simeq 1.07-1.14$~\cite{barbosa},
$E_i > E_{\rm thr}$. Thus, replacing $E_i$ by $E_{\rm thr}$,
\begin{eqnarray}
\nonumber f_\gamma^{\rm det}(E\ge E_{\rm thr})& > &
\frac{
\int_{E_{\rm thr}} A_\gamma (E) \Phi_\gamma (E) dE
}{
\int_{E_{\rm thr}} A_\gamma (E) \Phi_\gamma (E) dE
+\sum_i\int_{E_{\rm thr}} A_i (E) \Phi_i (E) dE
}\\
& \ge &
\frac{
\int_{E_{\rm thr}} A_\gamma^{\rm min}  \Phi_\gamma (E) dE
}{
\int_{E_{\rm thr}} A_\gamma^{\rm min} \Phi_\gamma (E) dE
+\sum_i\int_{E_{\rm thr}} A_i (E) \Phi_i (E) dE
} ~,
\end{eqnarray}

where $A_\gamma^{\rm min}$ refers to the minimum value of
$A_\gamma (E\ge E_{\rm thr})$ and using 
$a/(a+b) \ge a^\prime/(a^\prime +b)$ for $a \ge a^\prime \ge 0$
and $b>0$.

Next, the acceptance ratio
$\epsilon_i (E) = A_\gamma^{\rm min} / A_i (E)$ is introduced,

\begin{equation}
f_\gamma^{\rm det}(E\ge E_{\rm thr}) >
\frac{
\int_{E_{\rm thr}} A_\gamma^{\rm min}  \Phi_\gamma (E) dE
}{
\int_{E_{\rm thr}} A_\gamma^{\rm min} \Phi_\gamma (E) dE
+\sum_i\int_{ E_{\rm thr}} \frac{A_\gamma^{\rm min}}{\epsilon_i (E)} \Phi_i (E)dE
} ~.
\end{equation}

From Fig.~\ref{fig-eff} the minimum acceptance ratio
$\epsilon_{\rm min} (E_{\rm thr}) \le \epsilon_i (E\ge E_{\rm thr}) $
can be extracted for each threshold energy $E_{\rm thr}$. 
In the current analysis,
$\epsilon_{\rm min} (E_{\rm thr}) \equiv
 \epsilon_{\rm fvc} (E_{\rm thr}) \simeq 0.72, 0.77, 0.77, 0.77$
for $E_{\rm thr}$ = 2, 3, 5, 10~EeV. Hence, it follows:

\begin{eqnarray}
\nonumber f_\gamma^{\rm det}(E\ge E_{\rm thr}) & > &
\frac{\int_{E_{\rm thr}} \Phi_\gamma (E) dE
}{
\int_{E_{\rm thr}}  \Phi_\gamma (E) dE
+ \frac{1}{\epsilon_{\rm fvc}(E_{\rm thr})} \sum_i\int_{E_{\rm thr}} \Phi_i (E) dE} \\
 & > & \epsilon_{\rm fvc}(E_{\rm thr}) \cdot
\frac{\int_{E_{\rm thr}} \Phi_\gamma (E) dE
}{
\int_{E_{\rm thr}}  \Phi_\gamma (E) dE
+ \sum_i\int_{E_{\rm thr}} \Phi_i (E) dE
}\\
\nonumber & = & \epsilon_{\rm fvc}(E_{\rm thr}) \cdot f_\gamma(E\ge E_{\rm thr})~,
\end{eqnarray}
where it was used that $\frac{1}{\epsilon_{\rm fvc}(E_{\rm thr})} > 1$.

Consequently, an upper limit $F_\gamma^{\rm ul}$ to the
fraction of photons in the cosmic-ray flux can conservatively
be calculated as
\begin{equation}
F_\gamma^{\rm ul} =
f_\gamma^{\rm ul,det} / \epsilon_{\rm fvc}
 > f_\gamma^{\rm det} / \epsilon_{\rm fvc} > f_\gamma ~.
\end{equation}

The upper limit obtained this way does not depend on 
assumptions about the differential fluxes $\Phi_\gamma (E)$
and $\Phi_i (E)$.


\end{document}